\documentclass[conference]{IEEEtran}
\IEEEoverridecommandlockouts
\usepackage{cite}
\usepackage{amsmath,amssymb,amsfonts}
\usepackage{algorithmic}
\usepackage{graphicx}
\usepackage{textcomp}
\usepackage{xcolor}
\usepackage{multirow}
\usepackage{tabu}
\usepackage{makecell}
\usepackage{color}
\usepackage{subfigure}
\usepackage{bm}
\usepackage{balance}

\def\BibTeX{{\rm B\kern-.05em{\sc i\kern-.025em b}\kern-.08em
    T\kern-.1667em\lower.7ex\hbox{E}\kern-.125emX}}
    
\def\BibTeX{{\rm B\kern-.05em{\sc i\kern-.025em b}\kern-.08em T\kern-.1667em\lower.7ex\hbox{E}\kern-.125emX}}

\begin{document}

\title{A Transfer Learning and Optimized CNN Based Intrusion Detection System for Internet of Vehicles
}

\author{\IEEEauthorblockN{Li Yang and Abdallah Shami}
\IEEEauthorblockA{ Western University, London, Ontario, Canada \\
e-mails: \{lyang339, abdallah.shami\}@uwo.ca}
}

\maketitle

\begin{abstract}
Modern vehicles, including autonomous vehicles and connected vehicles,  are increasingly connected to the external world, which enables various functionalities and services. However, the improving connectivity also increases the attack surfaces of the Internet of Vehicles (IoV), causing its vulnerabilities to cyber-threats. Due to the lack of authentication and encryption procedures in vehicular networks, Intrusion Detection Systems (IDSs) are essential approaches to protect modern vehicle systems from network attacks. In this paper, a transfer learning and ensemble learning-based IDS is proposed for IoV systems using convolutional neural networks (CNNs) and hyper-parameter optimization techniques. In the experiments, the proposed IDS has demonstrated over 99.25\% detection rates and F1-scores on two well-known public benchmark IoV security datasets: the Car-Hacking dataset and the CICIDS2017 dataset. This shows the effectiveness of the proposed IDS for cyber-attack detection in both intra-vehicle and external vehicular networks.

\end{abstract}

\section{Introduction}
With the rapid development of the Internet of Things (IoT) and the Internet of Vehicles (IoV) technologies, modern vehicles have been evolving to network-controlled vehicles, including Autonomous Vehicles (AVs) and Connected Vehicles (CVs) \cite{DTL1}. Typical IoV systems involve intra-vehicle networks (IVNs) and external networks. In IVNs, Controller Area Network (CAN) bus is the central system that enables communications between Electronic Control Units (ECUs) to perform actions and adopt functionalities. On the other hand, external vehicular networks allow the connection between smart vehicles and other entities in IoV, including road-side units, infrastructures, and road users \cite{mth}. 

However, the improving connectivity and accessibility of vehicular networks has increased the cyber-attack surfaces of modern vehicles \cite{tree}. Additionally, due to the limited length of CAN packets, there is no authentication or encryption strategies involved in the processing of these packets \cite{DTL1}. The lack of fundamental security measures enables cyber-attackers to inject malicious messages to IVNs and launch different types of attacks, like Denial of Service (DoS), fuzzy, and spoofing attacks. On the other hand, the emerging cellular connections between connected vehicles and external networks have made these vehicles vulnerable to various conventional cyber-attacks \cite{iovcic1}. Therefore, it is crucial to develop Intrusion Detection Systems (IDSs) to protect IoV systems and smart vehicles by identifying cyber-attacks \cite{CNN1}.

Recently, due to the progression of Machine Learning (ML) and Deep Learning (DL) techniques, their applications in cyber-security and vehicle systems have attracted the attention of researchers and automotive manufacturers \cite{tnsm} \cite{dcnn}. ML and DL techniques are widely used to develop classifier-based IDSs that can distinguish between normal network traffic and different cyber-attacks through traffic data analytics \cite{pwpae}. In this paper, an intelligent IDS model based on optimized Convolutional Neural Networks (CNNs), transfer learning, and ensemble learning techniques is proposed to protect IoV systems. Five advanced CNN models, including VGG16, VGG19, Xception, Inception, and InceptionResnet \cite{CNNs}, are used to train base learners on vehicle network traffic data. The hyper-parameters of the CNN models are tuned using Particle Swarm Optimization (PSO), a hyper-parameter optimization (HPO) method, to obtain optimized learning models \cite{hpme}. The base CNN models are then integrated using two ensemble strategies, confidence averaging and concatenation, to further improve the intrusion detection performance. The effectiveness and efficiency of the proposed IDS framework are evaluated using two public vehicle network datasets: the Car-Hacking dataset \cite{GIDS} and the CICIDS2017 dataset \cite{CIC}.

This paper mainly makes the following contributions:
\begin{enumerate}
\item It proposes a novel framework for effective cyber-attack detection in both intra-vehicle and external networks through CNN, transfer learning, ensemble learning, and HPO techniques\footnote{
code is available at: https://github.com/Western-OC2-Lab/Intrusion-Detection-System-Using-CNN-and-Transfer-Learning}. 
\item It proposes a data transformation method that can effectively transform vehicle network traffic data into images to more easily distinguish various cyber-attack patterns.
\item It evaluates the proposed method on two benchmark cyber-security datasets that represent intra-vehicle and external network data, and compares the model's performance with other state-of-the-art methods.
\end{enumerate}


To the best of our knowledge, no previous work proposed such an optimized IDS model that integrates CNN, transfer learning, ensemble learning, and HPO techniques to effectively detect various types of attacks on both intra-vehicle and external networks.

The rest of the paper is organized as follows. Section II introduces the related work that uses ML and DL algorithms for vehicle network intrusion detection. Section III presents the proposed framework, including data transformation, CNN, transfer learning, ensemble learning, and HPO techniques. Section IV presents and discusses the experimental results. Finally, Section V summarizes the paper.

\section{Related Work}
ML and DL models have been widely used in IoV intrusion detection tasks. Rosay \textit{et al.} \cite{iovcic1} proposed a DL-based IDS for connected vehicles using Multi-Layer Perceptron (MLP). The MLP model was evaluated on an automotive microprocessor using the CICIDS2017 dataset. Yang \textit{et al.} \cite{mth} \cite{tree} proposed a tree-based stacking algorithm for network traffic analysis in IoV environments. The proposed stacking method shows high performance on the IoV and CICIDS2017 datasets. 

Several existing works focused on CNN-based IDS development for vehicular networks. Mehedi \textit{et al.} \cite{DTL1} proposed the P-LeNet method for in-vehicle network intrusion detection based on deep transfer learning. The P-LeNet model achieved a high F1-score of 97.83\% on the Car-Hacking dataset. Hossain \textit{et al.} \cite{CNN1} proposed a one-dimensional CNN (1D-CNN) based IDS for intra-vehicle intrusion detection, as 1D-CNN models work well in many time-series data analytics problems. Song \textit{et al.} \cite{dcnn} proposed a deep CNN (DCNN) based IDS model using reduced InceptionResnet to detect attacks in IVNs. The DCNN model shows high accuracy on the Car-Hacking dataset. 

Although the above methods achieve high accuracy in IoV cyber-attack detection tasks, there is still much room for performance improvement. The proposed solution aims to construct an optimal IDS framework using state-of-the-art CNN models optimized using HPO and ensemble learning strategies. Additionally, transfer learning techniques are used to improve the model training efficiency.

\section{Proposed Framework}
\subsection{System Overview}
\begin{figure}
     \centering
     \includegraphics[width=8.5cm]{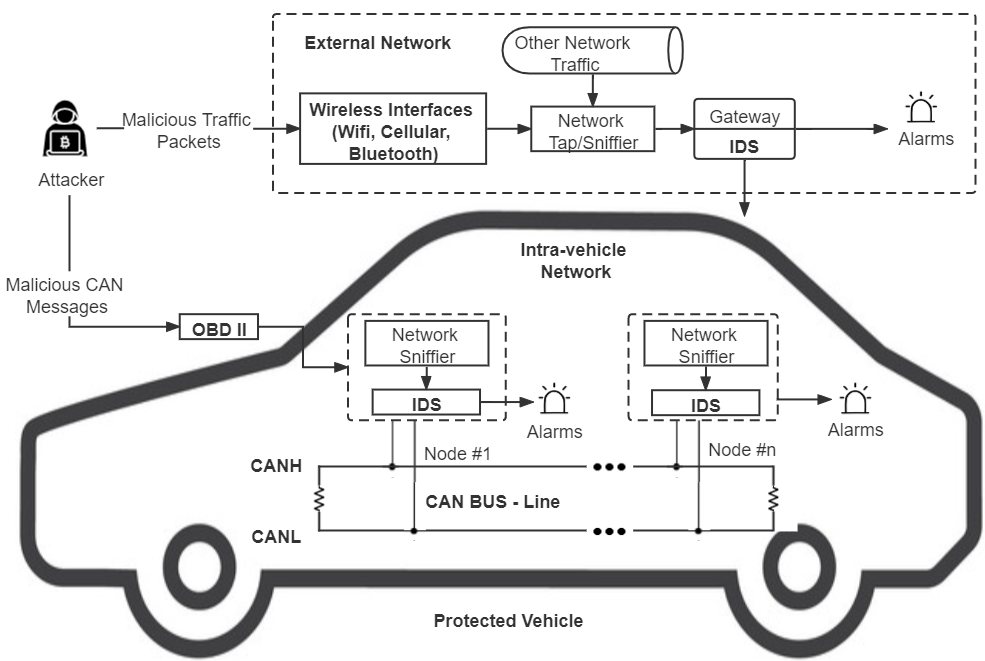}
     \caption{The IDS-protected vehicle architecture.} 
     \label{deploy}
\end{figure}

The purpose of this work is to develop an IDS that can detect various types of attacks in intra-vehicle and external vehicular networks to protect them both. The typical attack scenario and the architecture of an IDS-protected vehicle are shown in Fig. \ref{deploy}. Cyber-attackers can launch internal attacks on IVNs through the On-Board Diagnostics II (OBD II) interface and launch external attacks to external vehicular networks through wireless interfaces by sending malicious traffic packets. Thus, the proposed IDS should be deployed in both IVNs and external networks. In IVNs, the proposed IDS can be deployed on top of the CAN-bus to detect abnormal CAN messages and generate alarms \cite{tree}. In external networks, the proposed IDS can be incorporated into the gateways to identify and block all malicious packets that aim to breach the vehicles \cite{mth}. 

\begin{figure}
     \centering
     \includegraphics[width=7.3cm]{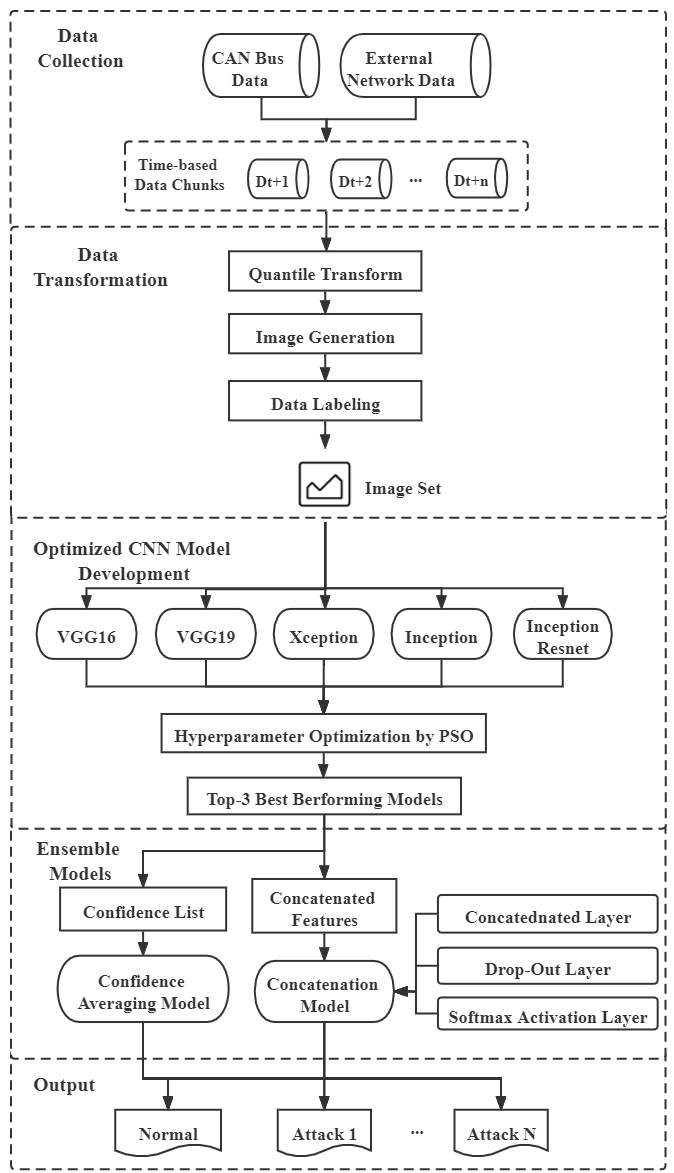}
     \caption{The proposed optimized CNN-based IDS framework.} \label{frame}
\end{figure}

In this paper, a novel optimized CNN and transfer learning-based IDS is proposed to detect various types of attacks in IoV systems. Figure \ref{frame} demonstrates the overview of the proposed IDS framework. Firstly, the intra-vehicle and external network data are collected in time-based chunks and then transformed into images using the quantile transform method. At the next stage, the generated image set is trained by five state-of-the-art CNN models (VGG16, VGG19, Xception, Inception, and InceptionResnet) to construct base learners. The CNN models are optimized by PSO, a HPO method that can automatically tune the hyper-parameters. After that, the top-3 best performing CNN models are selected as the three base CNN models to construct the ensemble learning models. Lastly, two ensemble strategies, confidence averaging and concatenation, are used to construct ensemble models for final detection. 

\subsection{Data Description and Transformation}
To develop the proposed IDS for both IVNs and external vehicular networks, two datasets are used in this work. The first dataset is the Car-Hacking dataset \cite{GIDS} that represents intra-vehicle data, as it is generated by transmitting CAN packets into the CAN-bus of a real vehicle. The CAN identifier (ID) and 8-bit data field of CAN packets (DATA[0]-DATA[7]) are the main features of the dataset. The Car-Hacking dataset involves four main types of attacks: DoS, fuzzy, gear spoofing, and Revolutions Per Minute (RPM) spoofing attacks. The second dataset used is the CICIDS2017 dataset \cite{CIC} that represents external network data, as it is a state-of-the-art network security dataset that includes the most updated attack patterns. According to the dataset analysis in \cite{tree} \cite{cicdetailed}, the attack patterns in the CICIDS2017 dataset can be summarized into five main types of attacks: DoS attacks, port-scan attacks, brute-force attacks, web-attacks, and botnets. 

After acquiring the data, it should be pre-processed to generate a proper input for the proposed IDS. As CNN models work better on image sets and vehicular network traffic datasets are usually tabular data, the original network data should be transformed into image forms \cite{CNN2}.  

The data transformation process starts with data normalization. Since the pixel values of images range from 0 to 255, the network data should also be normalized into the scale of 0-255. Among the normalization techniques, min-max and quantile normalization are the two commonly used methods that can convert data values to the same range. As min-max normalization does not handle outliers well and may cause most data samples to have extremely small values, quantile normalization is used in the proposed framework \cite{norm}. The quantile normalization method transforms the feature distribution to a normal distribution and re-calculates all the features values based on the normal distribution. Therefore, the majority of variable values are close to the median values, which is effective in handling outliers \cite{norm}. 

After data normalization, the data samples are converted into chunks based on the timestamps and feature sizes of network traffic datasets. For the Car-Hacking dataset, as it has 9 important features (CAN ID and DATA[0]-DATA[7]), each chunk of 27 consecutive samples with 9 features ($27\times9=243$ feature values in total) are transformed into an image of shape $9\times9\times3$ \cite{CNN2}. Thus, each transformed image is a square color image with three channels (red, green, and blue). Similarly, the CICIDS2017 dataset with 20 important features generated from \cite{mth} is transformed to $20\times20\times3$ color images, so each chunk of this dataset consists of $20\times3=60$ consecutive data samples. As the images are generated based on the timestamps of the data samples, the time-series correlations of the original network data can be retained. 

In the next step, the transformed images are labeled based on the attack patterns in the data chunks. If all the samples in a chunk/image are normal samples, this image is labeled “Normal”. On the other hand, if a chunk/image contains attack samples, this image is labeled as the most frequent attack type in this chunk. For example, if a DoS attack occurs in a chunk with the highest proportion, the corresponding image will be labeled “DoS attack”.  

\begin{figure}
  \centering
  \subfigure[Car-Hacking dataset.]{
    \label{fig:subfig:a} 
    \includegraphics[width=4.05cm]{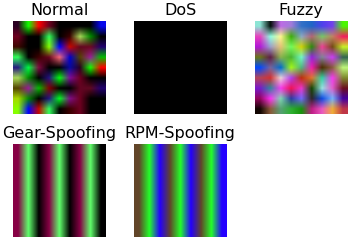}}
  \hspace{0.2cm}
  \subfigure[CICIDS2017.]{
    \label{fig:subfig:b} 
    \includegraphics[width=4.05cm]{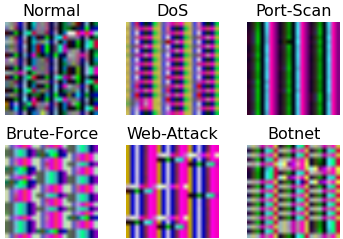}}
  \caption{The representative sample images of each class in two datasets: a) Car-Hacking dataset; b) CICIDS2017.}
    \label{fig:subfig} 
 \end{figure}

After the above data pre-processing procedures, the final transformed image set is generated as the input of CNN models. The representative samples for each type of attack in the Car-Hacking dataset and the CICIDS2017 dataset are shown in Fig. \ref{fig:subfig}. For the Car-Hacking dataset, it can be seen from Fig. \ref{fig:subfig:a} that there are large differences in the feature patterns between the normal samples and different types of attacks. The feature patterns of fuzzy attack images are more random than normal images, while DoS attack samples are high-frequency empty messages, causing pure black patterns. Gear and RPM spoofing attacks are launched by injecting messages with certain CAN IDs and packets to masquerade as legitimate users, so their images also have certain feature patterns \cite{GIDS}. Similarly, the attack patterns of CICIDS2017 can be obviously distinguished according to the feature patterns shown in Fig. \ref{fig:subfig:b}.

\subsection{CNN and Transfer Learning}
CNN is a common DL model that is widely used in image classification and recognition problems \cite{dcnn}. The images can be directly inputted into CNN models without additional feature extraction and data reconstruction processes. A typical CNN comprises three types of layers: convolutional layers, pooling layers, and fully-connected layers \cite{dcnn}. In convolutional layers, the feature patterns of images can be automatically extracted by convolution operations. In pooling layers, the data complexity can be reduced without losing important information through local correlations to avoid over-fitting. Fully-connected layers serve as a conduit to connect all features and generate the output.

For DL models, Transfer Learning (TL) is the process of transferring the weights of a Deep Neural Network (DNN) model trained on one dataset to another dataset \cite{TL}. The TL technique has been successfully applied to many image processing tasks. This is because the feature patterns learned by the bottom layers of CNN models are usually general patterns that are applicable to many different tasks, and only the features learned by the top layers are specific features for a particular dataset \cite{TL}. Therefore, the bottom layers of CNN models can be directly transferred to different tasks. To improve the effectiveness of TL, fine-tuning can be used in the TL process of DL models. In fine-tuning, most of the layers of the pre-trained model are frozen (\textit{i.e.}, their weights are retained), while a few of the top layers are unfrozen to re-train the model on a new dataset. Fine-tuning enables the learning model to update the higher-order features in the pre-trained model to better fit the target task or dataset \cite{TL}.

In the proposed framework, we have selected VGG16, VGG19, Xception, Inception, and InceptionResnet as the base CNN models due to their success in most image classification problems \cite{CNNs}. These CNN models are pre-trained on the ImageNet dataset and have demonstrated great performance on general image classification tasks. The ImageNet dataset is a benchmark image processing dataset that has more than one million images of 1,000 classes \cite{CNNs}.

The VGG16 models with 16 layers (VGG16) and with 19 layers (VGG19) proposed in \cite{vgg} have achieved a reduced error rate of 7.3\% on the ImageNet Challenge. The VGG16 architecture comprises five blocks of convolutional layers and three fully connected layers, while the VGG19 architecture has three more convolutional layers. The Inception network introduced in \cite{inception} uses convolutional feature extractors that combine different contexts to obtain different types of feature patterns, which reduces the computational cost through dimensionality reduction. Xception \cite{xception} is an extension of the Inception network that uses depthwise separable convolutions to replace the standard network convolutions. The memory requirement of Xception is slightly smaller than Inception. InceptionResnet is another extension of Inception that incorporates the residual connections from Resnet into the Inception network \cite{CNNs}. InceptionResnet outperforms Inception models on image classification challenges, but it requires twice the computational operations and memory than Inception. 

After using transfer learning and fine-tuning to train five state-of-the-art CNN models on the vehicle network datasets, the top-3 best performing CNN models are selected as the base learners to construct the ensemble models that are introduced in the next subsection.

\subsection{Proposed Ensemble Learning Models}
Ensemble learning is a technique that integrates multiple base learning models to construct an ensemble model with improved performance. Ensemble learning is widely used in data analytics problems because an ensemble of multiple learners usually performs better than single learners \cite{mth}. 

Confidence averaging is an ensemble learning approach that combines the classification probability values of base learners to find the class with the highest confidence value \cite{ensemble}. In DL models, softmax layers can output a posterior probability list that contains the classification confidence of each class. The confidence averaging method calculates the average classification probability of base learners for each class, and then returns the class label with the highest average confidence value as the final classification result. The confidence value of each class is calculated using the softmax function \cite{ensemble}:   

\begin{equation}
\operatorname{Softmax}(\mathbf{z})_{i}=\frac{e^{\mathbf{z}_{i}}}{\sum_{j=1}^{C} e^{\mathbf{z}_{j}}}
\end{equation}
Where $\mathbf{z}$ is the input vector, $C$ is the number of classes in the dataset, $e^{\mathbf{z}_{i}}$ and $e^{\mathbf{z}_{j}}$ are the standard exponential functions for the input and output vectors, respectively.

The predicted class label obtained by the confidence averaging method can be denoted by:
\begin{equation}
\hat{y}=\underset{i \in\{1, \cdots, c\}}{\operatorname{argmax}} \frac{\sum_{j=1}^{k} p_{j}\left(y=i \mid B_{j}, \boldsymbol{x}\right)}{k}
\end{equation}

Where $B_{j}$ is the $j_{th}$ base learner, $k$ is the number of selected base CNN learners, and $k=3$ in the proposed IDS; $p_{j}(y=i \mid B_{j}, \boldsymbol{x})$ indicates the prediction confidence of a class value $i$ in a data sample $\boldsymbol{x}$ using $B_j$.
	
Unlike the conventional voting method that only considers the class labels, confidence averaging enables the ensemble model to detect uncertain classification results and correct the misclassified samples through the use of classification confidence. The computational complexity of an entire ensemble model depends on the complexity of base learners, while the time complexity for the confidence averaging method itself is only $O(NKC)$, where $N$ is the number of instances, $K$ is the number of base CNN models, and $C$ is the number of classes \cite{oasw}. As $K$ and $C$ are usually small, the execution speed of the confidence averaging method is usually high.
	
Concatenation \cite{conca} is another ensemble strategy for DL models. A concatenated CNN aims to extract the highest order features generated from the top dense layer of base CNN models and use concatenate operations to integrate all the features into a new concatenated layer that contains all the features. The concatenated layer is followed by a drop-out layer to reduce redundant features and a softmax layer to construct a new CNN model. The advantage of concatenation is that it can combine the highest order features to construct a comprehensive new model. However, as the new model needs to be re-trained on the entire dataset, it introduces additional model training time. The computational complexity of the concatenation method is $O(NF)$, where $N$ is the number of data samples, and $F$ is the total number of features extracted from the dense layers of the base CNN models. 

\subsection{Hyper-Parameter Optimization (HPO)}
To better fit the base models to the selected datasets and further improve the models’ performance, the hyper-parameters of CNN models need to be tuned and optimized 

Similar to other DL models, CNN models have a large number of hyper-parameters that need tuning. These hyper-parameters can be classified as model-design hyper-parameters and model-training hyper-parameters \cite{hpme}. Model-design hyper-parameters are the hyper-parameters that should be set in the model design process. In the proposed TL framework, the model-design hyper-parameters include the number or percentage of frozen layers, the learning rate, and the drop-out rate. On the other hand, model-training hyper-parameters are used to balance the training speed and model performance, involving the batch size, the number of epochs, and early stop patience. The above hyper-parameters have a direct impact on the structure, effectiveness, and efficiency of CNN models. 

HPO is an automated process of tuning hyper-parameters of ML or DL models using optimization techniques \cite{hpme}. Among the optimization techniques used for HPO problems, PSO is a widely-used metaheuristic optimization method that identifies optimal hyper-parameter values via the information sharing and cooperation among the particles in a swarm \cite{hpme}. At the initial stage of PSO, each individual in the group is initialized with a position $\overrightarrow{x_{i}}$ and velocity $\overrightarrow{v_{i}}$. After each iteration, the velocity of each particle are updated based on its own current best position $\overrightarrow{p_{i}}$ and the current global optimal position $\overrightarrow{p}$ shared by other individuals: 
\begin{equation}
\overrightarrow{v_{i}}:=\overrightarrow{v_{i}}+U\left(0, \varphi_{1}\right)(\overrightarrow{p_{i}}-\overrightarrow{x_{i}})+U\left(0, \varphi_{2}\right)(\overrightarrow{p}-\overrightarrow{x_{i}}),
\end{equation}
where $U(0, \varphi)$ is the continuous uniform distribution calculated by the acceleration constants $\varphi_{1}$ and $\varphi_{2}$. 

Finally, the particles can gradually move towards the promising regions to identify the global optimum. PSO is chosen in the proposed framework due to its support to different types of hyper-parameters and its low time complexity of $O(NlogN)$ \cite{hpme}.

\section{Performance Evaluation}
\subsection{Experimental Setup}
The experiments were conducted using Scikit-learn and Keras libraries in Python. In the experiments, the proposed DL models were trained on a Dell Precision 3630 machine with an i7-8700 processor and 16 GB of memory and tested on a Raspberry Pi 3 machine with a BCM2837B0 64-bit CPU and 1 GB of memory, representing an IoV central server machine and a vehicle-level local machine, respectively.

The proposed framework is evaluated on two benchmark vehicle network security datasets, Car-Hacking \cite{GIDS} and CICIDS2017 \cite{CIC} datasets, as described in Section III-B. Five-fold cross-validation is used to evaluate the proposed model, which can avoid over-fitting and biased results. On the other hand, as network traffic data is usually highly imbalanced data that only has a small percentage of attack samples, four different metrics, including accuracy, precision, recall, and F1-scores, are used for performance evaluation. Furthermore, to evaluate the efficiency of the proposed method, model training time on the server-level machine and model testing time on the vehicle-level machine are also monitored and compared.

\subsection{Experimental Results and Discussion}
To construct optimal models, the major hyper-parameters of all the base CNN models in the proposed framework were optimized using PSO. As CNN models with default hyper-parameter values can already achieve near 100\% accuracy on the Car-Hacking dataset, the HPO process was only implemented for the CICIDS2017 dataset. Table \ref{hpo} illustrates the initial search range and the optimal values of the hyper-parameters. After HPO, the optimized CNN models were used as base learners to construct the proposed ensemble models.

\begin{table}[]
\caption{Hyper-parameter Configuration of CNN models}
\centering
\setlength\extrarowheight{1pt}
\scalebox{0.82}{
\begin{tabular}{|>{\centering\arraybackslash}p{8em}|>{\centering\arraybackslash}p{8em}|>{\centering\arraybackslash}p{7em}|>{\centering\arraybackslash}p{7em}|}
\hline
\textbf{Hyper-Parameter}                 & \textbf{Model}                            & \textbf{Search Range} & \textbf{Optimal Value} \\ \hline
Number of epochs                         & \multirow{5}{*}{\shortstack{General (All \\CNN models)}} & [5, 50]               & 20                     \\ \cline{1-1} \cline{3-4} 
Batch size                               &                                           & [32, 128]             & 128                    \\ \cline{1-1} \cline{3-4} 
Early stop patience                      &                                           & [2, 5]                & 3                      \\ \cline{1-1} \cline{3-4} 
Learning rate                            &                                           & (0.001, 0.1)          & 0.003                  \\ \cline{1-1} \cline{3-4} 
Drop-out rate                            &                                           & (0.2, 0.8)            & 0.5                    \\ \hline
\multirow{6}{*}{\shortstack{Number of frozen\\ layers}} & Xception                                  & [60,125]              & 121                    \\ \cline{2-4} 
                                         & Vgg16                                     & [8,   16]             & 15                     \\ \cline{2-4} 
                                         & Vgg19                                     & [10,   19]            & 19                     \\ \cline{2-4} 
                                         & Inception                               & [80,   159]           & 148                    \\ \cline{2-4} 
                                         & InceptionResnet                         & [300,   572]          & 522                    \\ \hline
\end{tabular}
}
\label{hpo}%
\end{table}

\begin{table}[]
\caption{Performance Evaluation of Models on Car-hacking Dataset}
\centering
\setlength\extrarowheight{1pt}
\scalebox{0.82}{
\begin{tabular}{|>{\centering\arraybackslash}p{6em}|>{\centering\arraybackslash}p{3.7em}|>{\centering\arraybackslash}p{4.0em}|>{\centering\arraybackslash}p{2.6em}|>{\centering\arraybackslash}p{2.4em}|>{\centering\arraybackslash}p{3.8em}|>{\centering\arraybackslash}p{4.1em}|}
\hline
\textbf{Method}                         & \textbf{Accuracy (\%)} &\textbf{Precision (\%)} & \textbf{Recall (\%)} &\textbf{F1 (\%)}  &\textbf{Training Time (s)} & \textbf{Test Time Per Packet (ms)} \\ \hline
P-LeNet \cite{DTL1}                         & 98.10        & 98.14         & 98.04      & 97.83   & -                 & -                            \\ \hline
1D-CNN \cite{CNN1}                         & 99.96        & 99.94         & 99.63      & 99.80   & -                 & -                            \\ \hline
DCNN \cite{dcnn}                           & 99.93        & 99.84         & 99.84      & 99.91   & -                 & -                            \\ \hline
VGG16-PSO                       & 99.97       & 99.97       & 99.97    & 99.97 & 384.9             & 0.2                          \\ \hline
VGG19-PSO                       & 100.0        & 100.0         & 100.0      & 100.0   & 417.9             & 0.2                          \\ \hline
Xception-PSO                    & 100.0        & 100.0         & 100.0      & 100.0   & 529.2             & 0.3                          \\ \hline
Inception-PSO                 & 100.0        & 100.0         & 100.0      & 100.0   & 733.6             & 0.6                          \\ \hline
InceptionResnet -PSO           & 100.0        & 100.0         & 100.0      & 100.0   & 970.4             & 1.3                          \\ \hline
\textbf{Concatenation (Proposed)}        & \textbf{100.0}        & \textbf{100.0 }        & \textbf{100.0}      & \textbf{100.0}   & \textbf{2490.5}            & \textbf{3.2}                          \\ \hline
\textbf{Confidence Averaging (Proposed)} & \textbf{100.0 }       & \textbf{100.0}         & \textbf{100.0 }     & \textbf{100.0 }  & \textbf{1680.7 }           & \textbf{2.7}                          \\ \hline
\end{tabular}
}
\label{r1}%
\end{table}

\begin{table}[]
\caption{Performance Evaluation of Models on CICIDS2017 Dataset}
\centering
\setlength\extrarowheight{1pt}
\scalebox{0.82}{
\begin{tabular}{|>{\centering\arraybackslash}p{6em}|>{\centering\arraybackslash}p{3.6em}|>{\centering\arraybackslash}p{3.8em}|>{\centering\arraybackslash}p{2.9em}|>{\centering\arraybackslash}p{2.6em}|>{\centering\arraybackslash}p{3.8em}|>{\centering\arraybackslash}p{4.1em}|}
\hline
\textbf{Method}                         & \textbf{Accuracy (\%)} &\textbf{Precision (\%)} & \textbf{Recall (\%)} &\textbf{F1 (\%)}  &\textbf{Training Time (s)} & \textbf{Test Time Per Packet (ms)} \\ \hline
KNN \cite{CIC}                      & 96.3  & 96.2  & 93.7   & 96.3  & 15243.6 & 0.2  \\ \hline
RF \cite{CIC}                      & 98.82 & 98.8  & 99.955 & 98.8  & 1848.3  & 0.3  \\ \hline
MLP \cite{iovcic1}                    & 99.46 & 99.52 & 99.40  & 99.46 & -       & 1.1 \\ \hline
VGG16-PSO                       & 99.724 & 99.625 & 99.724 & 99.674 & 436.5  & 0.1 \\ \hline
VGG19-PSO                       & 99.849 & 99.850 & 99.849 & 99.850 & 688.1  & 0.1 \\ \hline
Xception-PSO                    & 99.699 & 99.700 & 99.699 & 99.697 & 655.5  & 0.2 \\ \hline
Inception-PSO                 & 99.750 & 99.725 & 99.750 & 99.729 & 782.8  & 0.3 \\ \hline
InceptionResnet -PSO           & 99.849 & 99.850 & 99.849 & 99.850 & 1187.2 & 0.7 \\ \hline
\textbf{Concatenation (Proposed)}        &\textbf{99.899} & \textbf{99.900} & \textbf{99.899} & \textbf{99.898} & \textbf{3598.7} & \textbf{1.8} \\ \hline
\textbf{Confidence Averaging (Proposed)} & \textbf{99.925} & \textbf{99.925} & \textbf{99.924} & \textbf{99.925} &\textbf{2658.1} &\textbf{1.5} \\ \hline
\end{tabular}
}
\label{r2}%
\end{table}

The results of evaluating the optimized CNN models and the proposed ensemble models on the Car-Hacking and CICIDS2017 datasets are shown in Tables \ref{r1} \& \ref{r2}, respectively. As shown in Table \ref{r1}, all optimized base CNN models except VGG16 achieve 100\% accuracy and F1-scores. This is mainly because the normal and attack patterns in the Car-Hacking dataset can be obviously distinguished through the transformed images shown in Fig. 3. The two ensemble techniques, concatenation and confidence averaging methods, can also achieve 100\% F1-scores, while the total training time of confidence averaging is much lower than concatenation (1680.7 s versus 2490.5 s). Therefore, the confidence averaging method is more efficient. Moreover, the performance of the proposed models is compared with other state-of-the-art methods \cite{DTL1} \cite{CNN1} \cite{dcnn}. As shown in Table \ref{r1}, most of the compared approaches achieve high accuracy due to the simplicity of the Car-Hacking dataset. Among the models shown in Table \ref{r1}, the proposed ensemble methods show the best performance by achieving at least 0.09\% F1-score improvements.

For the CICIDS2017 dataset, the optimized base CNN models achieve high F1-scores of 99.674\% to 99.850\% after implementing data transformation and PSO, as shown in Table \ref{r2}. The proposed confidence averaging ensemble model also achieves the highest F1-score of 99.925\%, which is slightly higher than the F1-score of the concatenation model (99.899\%). The two ensemble models also outperform other recent methods in the literature \cite{iovcic1} \cite{CIC}. Additionally, the total training time of the confidence averaging is also much lower than the concatenation approach.

The higher performance of the proposed models when compared with other state-of-the-art IDSs supports the reasons for using CNN, TL, and HPO techniques. Furthermore, the average test/prediction time of the proposed ensemble models for each packet on the Raspberry Pi machine is at a low level, from 1.5 ms to 3.2 ms, as shown in Tables \ref{r1} \& \ref{r2}. As the real-time requirement of vehicle anomaly detection systems is usually 10 ms for the analysis of each packet \cite{V2XAM1}, the low prediction time of the proposed models indicates the feasibility of applying the proposed IDS to real-time IoV systems. 

\section{Conclusion}
As modern vehicles are increasingly connected, the cyber-threats to IoV systems are also increasing significantly. To protect connected vehicles from being breached by cyber-attacks, this work proposed a transfer learning and ensemble learning-based IDS framework that uses optimized CNN models to identify various types of attacks in IoV systems. Additionally, a chunk-based data transformation method is proposed to transform vehicle network traffic data to image data used as the input of CNN models. The proposed IDS is evaluated on the Car-Hacking and CICIDS2017 dataset, representing intra-vehicle and external network data, respectively. The experimental results show that the proposed IDS framework can effectively identify various types of attacks with higher F1-scores of 100\% and 99.925\% than other compared state-of-the-art methods on the two benchmark datasets. Moreover, the model testing results on a vehicle-level machine show the feasibility of the proposed IDS in real-time vehicle networks. In future work, this framework will be extended to develop an online adaptive model that can achieve online learning and address concept drift in time-series vehicle network data.

\end{document}